\documentclass[11pt,a4paper,oneside]{article}

\usepackage[T1]{fontenc}
\usepackage[english]{babel}

\usepackage{textcomp}
\usepackage[noBBpl]{mathpazo}
\usepackage{booktabs}

\usepackage[cmex10]{amsmath}

\usepackage{textcomp}
\usepackage[noBBpl]{mathpazo}
\usepackage{booktabs}
\usepackage{amsfonts}
\usepackage{amsthm}

\usepackage{graphicx}

\usepackage{algorithm}
\usepackage{algorithmic}

\usepackage{array}

\usepackage{cite}
\usepackage{hypernat}
\usepackage{hyperref}
\usepackage[all]{hypcap}

\renewcommand{\epsilon}{\varepsilon} 

\renewcommand{\tilde}{\widetilde} 

\def\1{\mbox{1\hspace{-.35em}1}}

\newcommand{\btheta}{\mbox{\boldmath$\theta$}}

\newcommand{\bomega}{\mbox{\boldmath$\omega$}}
\newcommand{\bgamma}{\mbox{\boldmath$\gamma$}}

\begin{document}

\setlength{\parskip}{\baselineskip}
\setlength{\parindent}{0pt}


\author{Ir\`ene Gannaz\\
~\\ 
{\em Universit\'e de Lyon}\\
{\em CNRS UMR 5208}\\
{\em INSA de Lyon}\\
{\em Institut Camille Jordan}\\
{\em 20, avenue Albert Einstein}\\
{\em 69621 Villeurbanne Cedex}\\
{\em France.}\\
{\em E-mail: \url{irene.gannaz@insa-lyon.fr}}\\
}

\title{Classification of EEG recordings in auditory brain activity via a logistic functional linear regression model}

\date{\small June 2014}


\maketitle

{{\bf Abstract.} We want to analyse EEG recordings in order to investigate the phonemic categorization at a very early stage of auditory processing. This problem can be modelled by a supervised classification of functional data. Discrimination is explored via a logistic functional linear model, using a wavelet representation of the data. Different procedures are investigated, based on penalized likelihood and principal component reduction or partial least squares reduction.
}

\section*{Introduction}

We are interested in the categorization phenomenon in auditory brain activity. We consider EEG recordings measuring the cerebral activity in response to auditory stimuli. We would like to determine whether the different signals contain information on the sound heard by the patient and if we are able to discriminate the different audio stimuli. To this objective we propose a logistic functional linear model to study the classification of EEG curves. 


In literature, recordings usually deal with brain-computer interface \cite{Lotte} or with different biological contexts such as epilepsy where the characteristics of EEG recordings are much more apparent than in the auditory activity, e.g. \cite{Subasi, Rajendra}. The specificity of our data lies in the fact that the discriminative characteristics of the signals are very localized and standard methodologies failed to extract the information.

As discriminative properties are expected to be localized in time and in frequency in the signals, we use a wavelet representation of the EEG recordings. To enforce a reduction of the dimension of the model we introduce a constraint of sparsity, possibly coupled with principal components or partial least square. We study empirically our procedure and compare it to similar concepts \cite{OgdenReiss13}.

\section{Problem formulation}

EEG signals measure the human perception of bilabial plosives, here {\it /b/} versus {\it /p/}. Stimuli correspond to sounds {\it /ba/}, {\it /pa/} and two intermediates obtained by modifying the voicing onset times of the plosives {\it /b/} and {\it /p/}, taking intermediate values. We refer to \cite{speech} for a detailed description of the experiment. Neuroscientist are interested in the phonemic categorization: when asking for the identification of the intermediate sounds as {\it /ba/} or {\it /pa/}, the first intermediate stimuli is usually recognized as a {\it /ba/} and the second one as a {\it /pa/}.

\begin{figure}[!h]
\begin{center}
\includegraphics[scale=0.3]{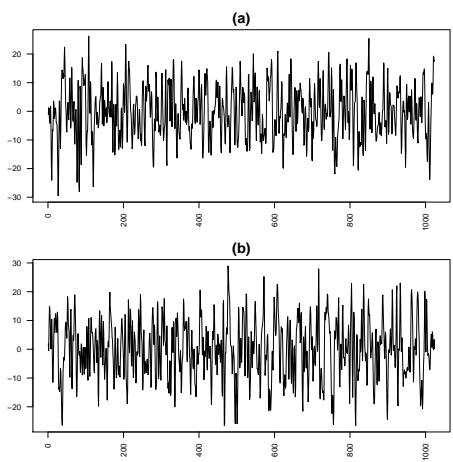}
\end{center}
\caption{Average of 10 EEG signals (a) with the stimulus {\it /ba/} and (b) with the stimulus {\it /pa/}.}
\label{fig:reponses}
\end{figure}

Only the evoked potentials between the frontal electrode and the right ear of the subject are considered. A high-pass filter at 80 Hertz was applied to keep only the frequencies corresponding to auditory activity. In addition, we consider averages of ten signals in order to get rid of a possible random effect. Two examples of resulting signals for pure sounds {\it /ba/} or {\it /pa/} are given in Figure~\ref{fig:reponses}. Note that EEG offers the possibility to combine many potentials on a single recording. A perspective of the present work is to take advantage of this multiplicity \cite{Rivet}.

The objective is two fold. We first aim to determine whether the recordings are informative. We moreover want to explore if the categorization is effective at this early stage of recording. Are the responses still discriminated in four classes or are they already categorized in two classes? Those questionings can be modelled by the prediction of a categorical variable $Y$ with respect to an explanatory functional variable $(X(t))_{t\in[0,1]}$. The label $Y$ corresponds to the stimuli and $X(\cdot)$ is the resulting EEG recording.

\section{The logistic functional linear regression}

Let $(Y_i,\{X_i(t),\,t\in[0,1]\})$, $i=1,\dots,n$, be independent observations. Responses $Y_i$ are labelled variables, with values $0$ or $1$. It can be generalized to more values but we focus on the Bernoulli case for brevity. Predictor variables $\left(\{X_i(t),\,t\in[0,1]\}\right)_{i=1,\dots,n}$ belong to the separable Hilbert space $L^2([0,1])$ with the usual inner product.

We consider a logistic functional linear model: \begin{equation}\label{eqn:modele}
\mathbb{P}\left(Y_i= 1 \,|\, X_i(\cdot)\right) = g(X_i(\cdot)),\end{equation}
$$ \text{~where~} g(X_i(\cdot))=\frac{e^{\eta\left(X_i(\cdot)\right)}}{1+e^{\eta\left(X_i(\cdot)\right)}}\\
\text{~and~} \eta\left(X_i(\cdot)\right)=<X_i(\cdot),\beta(\cdot)>.$$
The unknown function $\beta(\cdot)$ captures the features which discriminate the curves $(X_i(\cdot))_{i=1,\dots,n}$.

Logistic functional linear models received much interest last decade due to the large scope of applications. 
We are interested in this paper in an estimation scheme based on the decomposition of the explanatory curves on a given functional basis. This was investigated among others in \cite{James, MarxEilers, CardotSarda} with spline basis and different roughness penalties, and in \cite{ZhaoOgdenReiss} with wavelet basis. Authors in \cite{OgdenReiss13} propose to use both component reduction and roughness penalties, respectively with splines basis and with wavelet basis. The procedure proposed in this paper is based on similar concepts.

\section{Estimation procedure}

We introduce an orthogonal discrete wavelet transform $\mathcal W$ on $V_{j_0}\oplus_{j\geq j_0} W_{j}$, where $V_{j_0}$ is the space generated by the father wavelet at scale $j_0$ and $W_j$ is the space generated by the mother wavelet at scale $j$. 
 Every signal $X_i(\cdot)$ is decomposed by $\mathbf{X}_i=(X_i(t_j))_{j=1,\dots,d}=\mathcal W^T \btheta_i$, for $i=1,\dots,n$. The exponent $T$ denotes the transpose operator. The unknown function $\beta(\cdot)$ is also represented by the vector of its wavelet coefficients $\bomega$. Let $\{\omega_{\ell}\}_{\ell<2^{j_0}}$ be the scale coefficients while $\{\omega_{\ell}\}_{\ell>2^{j_0}}$ are the wavelet coefficients. The logistic functional linear regression model is expressed like a regression on the wavelet coefficients. 

As we are expecting that the discriminative function $\beta(\cdot)$ is localized in time and frequency, we impose the sparsity of the wavelet coefficients $\{\omega_\ell\}_{\ell>2^{j_0}}$. This is usually done in literature thanks to a $\ell^1$-penalization. This approach has been proposed by \cite{ZhaoOgdenReiss} in functional linear models with real responses. The authors establish the asymptotic consistency of the estimator. 

Following \cite{OgdenReiss13}, we can introduce an additional principal components reduction step to enforce the reduction of dimension. Let $a_1\geq a_2\geq \dots \geq a_d$ be the eigenvalues of $(\mathbf{\theta_i})_{i=1,\dots,n}$. We introduce the matrix $V_q$ of size $d\times q$ such as the $i$th column of the matrix $V_q$ is the eigenvector associated with the eigenvalue $a_i$. We then impose that $\bomega$ can be written $\bomega=V_q\bgamma$ with $\bgamma\in\mathbb{R}^q$. An extension to partial least squares reduction has also been explored.

The estimators are thus defined as follows:
\begin{eqnarray*}
\tilde{\bomega}_n(q,\lambda)&=&\mathop{argmin}_{\bomega}\, -\sum_{i=1}^n \mathcal L \left(Y_i,\,\theta_i^T \bomega\right)\,+\,\lambda\sum_{\ell=2^{j_0}+1}^d \left|\omega_\ell\right| \\
& & s.t.\,\,\, \exists \bgamma\in\mathbb{R}^q, \, \bomega=V_q\bgamma
\end{eqnarray*}
where $\mathcal L$ is the log-likelihood. The parameters $\lambda$ and $q$ are chosen by cross-validation. In literature, EEG classification is generally done applying a usual classification procedure for multivariate data on the wavelets coefficients or on statistics summarizing their behaviour (see e.g. \cite{Subasi, Rajendra}). The advantage of our procedure is to take into account the specificity of the wavelets coefficients, with the penalization enforcing sparsity. 

We compare our procedure with the estimators 
studied in \cite{OgdenReiss13}. Actually, we consider the following procedures, with a 5-folds cross validation:
\begin{itemize}
\item SPCR - A spline-based estimation, with Principal Component Reduction (CR) and a $\ell^2$ penalization. 
\item WNET - A wavelet-based estimation with a $\ell^1$-penalty.
\item WCR and WLS - A wavelet-based estimation, with respectively a sparse CR or a sparse Partial Least Squares (LS) reduction.
\item WPCR and WPLS - Our wavelet-based estimation described previously in the manuscript, with respectively CR or LS and a $\ell^1$-penalization.
\end{itemize}
We also implement A-WPCR and A-WPLS: the same as WPCR and WPLS but with parameters obtained by a corrected AIC procedure.

\section{Results}

We first try to discriminate the EEG recordings in four classes, with respect to the four stimuli. No estimators succeed in this classification. 
We then classify the signals in two classes, with {\it /ba/} and the first intermediate stimulus associated with a label 0 and {\it /pa/} and the second intermediate stimulus with a label 1.
Areas under ROC curves (AUC) are given in Table~\ref{tab:auc_4en2}. Following \cite{HosmerLemeshow}, we consider that the discrimination holds if the AUC is greater than 0.7. With the WNET estimator, the discrimination of the EEG signals in two classes is validated. Spline-based procedure is not able to capture the differences in signals. Estimators with a principal component or a partial least squares step succeed in discriminating on the learning sample but not on the validation sample. The reduction of component seems too dependent on the learning set and introduce a bias in the prediction power of the estimators. 


\begin{table}[!t]
\caption{AUC for discrimination of EEG recordings in two classes. The learning set and the validation set contain respectively 75 curves and 25 curves for each stimulus.}
\label{tab:auc_4en2}
\begin{center}
\begin{tabular}{@{} l@{\hspace{0.5cm}} lccccc @{}}
\hline
Method &SPCR&{\bf WNET}& WCR& WPCR& A-WPCR \\
\hline
Learning sample & 0.586  &{\bf 0.734}& 0.808 & 0.744 & 0.730 \\
Validation sample& 0.533 & {\bf 0.708}& 0.659 & 0.648  & 0.666\\  
\hline\\
\hline
Method &&& WLS& WPLS& A-WPLS \\
\hline
Learning sample & & &1 & 0.740 &1\\
Validation sample& & & 0.580 &0.572 & 0.597\\  
\hline
\end{tabular}
\end{center}
\end{table}

\begin{figure}[!h]
\begin{center}
\includegraphics[width=6cm,height=4cm]{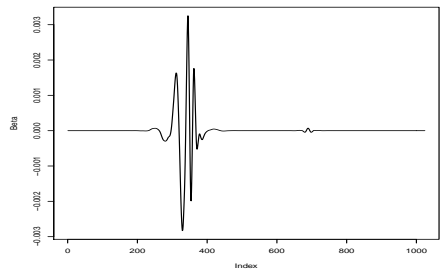}
\end{center}
\caption{Estimation of the discriminative function $\beta(\cdot)$ obtained by WNET when dicriminating EEG recordings in two classes.}
\label{fig:beta}
\end{figure}

The fact that the classification is effective establishes that recordings indeed contain information on the stimuli. We can moreover localize which part of the signals are informative to obtain this discrimination with the estimation of the function $\beta(\cdot)$. The estimation given by WNET in Figure~\ref{fig:beta} highlights the actual sparsity. 
 On the second hand, this result infers that the categorization of the plosives is probably done at an early stage of auditory activity. Yet, we are not able to reject the assumption of no-categorization, only to state that the categorization is more likely.

\section*{Conclusion}

This paper explores different estimation procedures in logistic functional linear models, based on a wavelet decomposition of the explanatory curves and principal component reduction. Our study stresses that this model with wavelet-based estimation seems efficient to discriminate complex signals such as EEGs but that the component reduction step introduce a bias. We identify a procedure that is able to extract where is the pertinent information contained in signals, in time and frequency. The application on signals relative to auditory activity highlights that the assumption of a categorization of sounds at an early stage of the auditory process seems likely.

\vspace{\baselineskip}

\noindent {\small {\bf Acknowledgements:} This work is part of a project supported by the IXXI research institute. It was motivated by a collaboration with Ludovic Bellier, Rafael Laboissi\`ere and Fabien Millioz, of the DyCog Team, in Lyon Neuroscience Research Center, France. The author is grateful to them for the data collection.}

\end{document}